\documentclass[11pt]{article}
\usepackage[final]{acl}

\usepackage{times}
\usepackage{latexsym}
\usepackage[T1]{fontenc}
\usepackage[utf8]{inputenc}
\usepackage{microtype}
\usepackage{inconsolata}

\usepackage{graphicx}
\usepackage{booktabs}
\usepackage{multirow}
\usepackage{subcaption}
\usepackage{amsmath}
\usepackage{amssymb}
\usepackage{xcolor}
\usepackage{url}
\usepackage{placeins}
\usepackage{stfloats}

\newcommand{\tightpara}[1]{%
  \par\noindent\textbf{#1}\hspace{0.5em}\ignorespaces}

\title{PULSAR: Pooled Unified Late-Interaction Search and Retrieval for Enterprise Visual Document RAG}

\author{%
  \textbf{Benjamin Constable\textsuperscript{1}} \quad
  \textbf{Anup Roy\textsuperscript{2}} \quad
  \textbf{Vishal Sharma\textsuperscript{2}} \\
  \textbf{Rishabh Gyanendra Upadhyay\textsuperscript{2}} \quad
  \textbf{Robin Mills\textsuperscript{2}} \quad
  \textbf{Aidan Philip Millar\textsuperscript{3}} \\[0.4em]
  \textsuperscript{1}Microsoft, London, UK \qquad
  \textsuperscript{2}Inception42, Abu Dhabi, UAE \\
  \textsuperscript{3}Mubadala Investment Company, Abu Dhabi, UAE%
}

\begin{document}
\maketitle
\raggedbottom

\begin{abstract}
Institutional investors search visually
dense pitch decks, board packs, and diligence materials that change hourly
near deal closing. OCR followed by figure verbalisation is costly to refresh
at this scale and can lose chart detail. We present PULSAR, a production
vision-first retrieval system deployed at Mubadala Investment Company.
PULSAR indexes page images with a frozen ColPali-style backbone and uses a
pooled two-stage late-interaction index: compact page summaries support
initial retrieval, followed by exact MaxSim rescoring over a finer pooled
representation. On ViDoRe V3, this design reduces median vector-search
latency by 15.1 times against an unpooled configuration with less than 0.01
absolute NDCG@10 and Recall@10 loss; production median vector-search latency is 156 ms.
Under concurrent load, the pooled index sustains approximately 88
times higher QPS than an unpooled index.
The event-driven ingestion path is estimated to be approximately 20 times
cheaper per page than the OCR+verbalisation baseline it replaced. Since March
2026, PULSAR has served 78 thousand documents and approximately 2.4 million pages
across more than 3,000 deals. At the production top K, it more than doubles
answer-fact recall over the OCR+verbalisation baseline.
\end{abstract}

\section{Introduction}
\label{sec:intro}

Enterprise RAG \cite{lewis2020rag} is harder than text-only
benchmarks \cite{thakur2021beir}: corporate decks and board
packs contain tables, legends, chart
annotations, and spatial cues that text-first pipelines lose even with
image verbalisation. Verbalisation is information-reducing. The standard pipeline applies OCR or layout-aware extraction
\cite{xu2020layoutlm,huang2022layoutlmv3}, chunks text, optionally
adds vision-language model (VLM) captions \cite{liu2023llava,hu2024docowl}, and retrieves with
hybrid search. Each page
is run through a document OCR+layout model and a per-figure VLM, then
the verbalised text is embedded: three passes where vision-first uses
one. At firm scale (${\sim}2.4$M chart-heavy pages), this is an order
of magnitude more work per page than a single embedding pass and is
repaid every time a deal document changes near closing.

Vision-first inverts this. The page image becomes the unit of indexing
and grounding. Ingestion collapses to a single forward pass of the
retriever's own embedding model per page, dropping the separate
OCR+layout and per-figure captioning passes and making cost scale with
page count alone. On enterprise corpora, \citet{loison2026vidore}
report visual retrievers outperforming
textual ones.

We build on VLM-based multi-vector page retrievers with
late-interaction scoring \cite{khattab2020colbert,faysse2024colpali},
rather than OCR-free document transformers
\cite{kim2022donut,hu2024docowl}. Late interaction keeps many embedding vectors per page (one per
image patch) rather than collapsing the page to a single vector, so
chart values and layout cues stay addressable. At firm scale this
puts billions of vectors in the index, and per-query
scoring cost grows linearly in patches per page.

PULSAR (Pooled Unified Late-interaction Search And Retrieval), our
production retrieval system, is built around this cost. It composes a
frozen ColPali-style backbone with a pooled two-stage retrieval
pipeline and an event-driven ingestion path. Our contribution is not
a new retrieval primitive but the engineering and empirical work of
scaling this vision-first paradigm to production at firm scale.

\tightpara{Contributions.}
(i) A decoupling of embedding-time and answer-time resolution:
150~DPI suffices for retrieval, while 500~DPI at answer time drives
page-legibility errors on firm decks to zero.
(ii) A two-stage pooled late-interaction index that combines mean
row/column pooling \cite{qdrant2026multivec_tut} for the first-stage
lookup with hierarchical-pooled exact rescoring
\cite{clavie2024tokenpool} and binary quantisation, selected by a
pooling-and-quantisation sweep. It cuts single-stream median vector
search latency $15.1\times$ at under 0.01 NDCG@10 and Recall@10 loss,
and holds a 156\,ms median under production concurrency.
(iii) An ingestion recipe (rolling rasterisation, strict-priority
work queues, page-hash resume) that scales with page count
rather than visual complexity. It is ${\approx}20\times$
cheaper per page than the OCR+verbalisation baseline it replaces
(call-count estimate, \S\ref{sec:ingest}).
(iv) A production deployment at Mubadala Investment Company (78k
documents across ${>}3{,}000$ deals, ${\sim}2.4$M pages, live since March~2026). An in-house
end-to-end answer evaluation shows PULSAR
exceeding an OCR+verbalisation baseline on context-fact recall,
answer-fact recall, and completeness at every top-$k$.

\section{Related Work}
\label{sec:related}

\tightpara{Vision-first document retrieval.} Late interaction began
with ColBERT \cite{khattab2020colbert,santhanam2022colbertv2}, which
gives each document one vector per token instead of a single dense
vector and scores a query--document pair by MaxSim: for each query
token, take its best match against any document token and sum. The
index grows, but matching becomes finer-grained. ColPali
\cite{faysse2024colpali} carries the idea to page images. A
vision-language backbone \cite{wang2024qwen2vl} splits the page into
patches, each patch becomes a token, and MaxSim then runs between
query text tokens and image patches.

\tightpara{Generate-and-encode hybrids.} The alternative to indexing
the image is to verbalise it first. Systems such as SERVAL
\cite{nguyen2025serval} verbalise page images before text-encoding
and themselves flag VLM-generated text as an unmeasured hallucination
source. Both paradigms run a VLM, but verbalisation bakes its output
into the index at ingestion; vision-first keeps the page image as the
stored unit and defers generation to answer time, grounded on the
retrieved page. Verbalisation often captures the concrete values
on a chart, but not how the items relate visually: the groupings and
orderings a reader uses to interpret it. Some of what analysts query
for is lost in that gap.

\tightpara{Generalisation to unseen corpora.} The verbalisation route
has one edge. \citet{most2025lostocr} report that OCR-based RAG
generalises better than vision-first retrieval \emph{fine-tuned on a
fixed corpus}. That finding applies to fine-tuning, which we avoid:
PULSAR uses a pretrained ColQwen3-4B with no firm-specific
fine-tuning, and still beats the baseline on the firm's own decks
(\S\ref{sec:headtohead}).

\tightpara{Vector indexing.} Dense retrieval leans on Hierarchical
Navigable Small World (HNSW) \cite{malkov2020hnsw} with
quantisation \cite{johnson2019billion} as its
approximate nearest neighbour (ANN) backend. Late interaction
strains that backend, since each page now carries many vectors
rather than one. Prior multivector engines answer with a
prefetch-then-rerank split: PLAID~\cite{santhanam2022plaid}
prefetches on centroids before a token-level rerank, and
EMVB~\cite{nardini2024emvb} compresses the prefetch to bit-vectors.
PULSAR keeps this split but combines two existing pooling methods
across its stages: mean row/column pooling~\cite{qdrant2026multivec_tut,yeroyan2026visualrag}
for the prefetch and hierarchical pooling~\cite{clavie2024tokenpool}
for the exact rerank (\S\ref{sec:retrieval}). Both are post-hoc and
need no retraining. ColMate~\cite{masry2025colmate} instead modifies training:
its TopKSim objective averages each query token's top-$K$ patch
similarities, reverting to plain MaxSim at inference. Recent
hybrid-vector work such as HEAVEN~\cite{kim2026heaven} uses a
single-vector first stage over visually summarised pages followed by
multi-vector reranking. PULSAR instead keeps multivectors at both
stages and serves them from a continuously updated production index.

\section{PULSAR System Design}
\label{sec:arch}
Late interaction buys finer-grained matching at a cost: each page
becomes many patch vectors rather than one, inflating both index size and
per-query scoring. PULSAR keeps this cost in check at each stage. A
single backbone pass (\S\ref{sec:backbone}) embeds each page at a
render DPI decoupled from the higher-resolution image kept for
answering (\S\ref{sec:dpi}), so cost scales with page count, not
visual complexity. Ingestion (\S\ref{sec:ingestion})
re-embeds only changed pages; retrieval (\S\ref{sec:retrieval}) scores
pooled summaries rather than full patch grids, keeping exact MaxSim
affordable.

\begin{figure*}[!t]
\centering
\includegraphics[width=\textwidth,keepaspectratio]{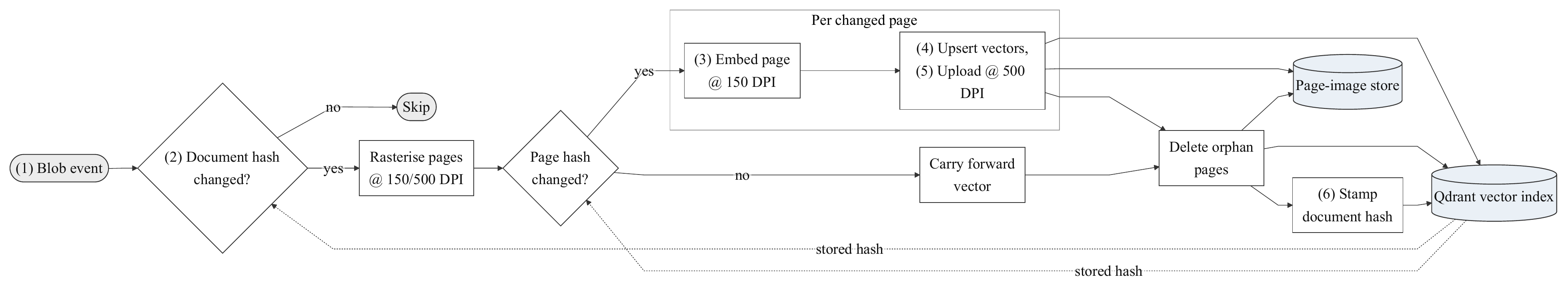}
\caption{PULSAR's ingestion path (\S\ref{sec:ingestion}): document- and
page-level hashes skip unchanged work, changed pages are re-embedded
with ColQwen3-4B and upserted, and the document hash is stamped once every page is
durable.}
\label{fig:ingest}
\end{figure*}

\subsection{Embedding backbone}
\label{sec:backbone}
PULSAR's embedding backbone is Tomoro ColQwen3-4B \cite{huang2025beyond}, a
ColPali-style multivector retriever on Qwen3-VL. At
deployment, it was the strongest Apache-2.0-licensed model within our
half-A10 serving budget (Table~\ref{tab:hw}); the 8B sibling scored
higher on ViDoRe~V3 \cite{loison2026vidore} but exceeded that budget.

\subsection{Architecture}
\label{sec:architecture}
The ingestion and retrieval paths of PULSAR\footnote{PULSAR builds on Microsoft's
open-source Multi-Modal RAG with ColPali accelerator
\cite{mmragcolpali} (vLLM serving, Helm/AKS deployment); the rest of
the system and the evaluation (\S\ref{sec:results}) are this paper's
contributions.} share two stores: a Qdrant vector index
and a page-image store. Figs.~\ref{fig:ingest}
and~\ref{fig:retrieval} show the corresponding components in execution
order.

\tightpara{Ingestion path.}
(1) A blob-storage event enters one of three priority queues.
(2) The dispatcher takes the next document; a document-level hash
skips it if unchanged, otherwise the indexer rasterises each page at
150~DPI for retrieval and 500~DPI for answer-time grounding, and a
page-level hash of the render marks which pages changed.
(3) ColQwen3-4B embeds the 150~DPI render of each changed page.
(4) The indexer derives mean-pooled row and column vectors for HNSW
prefetch and hierarchical-pooled vectors for exact reranking, then
upserts them into Qdrant.
(5) The 500~DPI render is written to the page-image store.
(6) The document hash is stamped only after every changed page has been
indexed and its page image persisted; queries filter out pages whose document has
not yet been stamped.

\tightpara{Retrieval path.}
(1) ColQwen3-4B embeds the text query once and leaves its token
multivectors unpooled.
(2) Two HNSW searches compare the query against the mean-pooled row
and column page representations.
(3) Their candidates jointly form a top-$k'$ shortlist.
(4) Exact MaxSim over the hierarchical-pooled page vectors reranks
that shortlist.
(5) The top-$k{=}5$ pages are returned.
(6) Their stored 500~DPI renders are fetched from the page-image store.
(7) The downstream VLM reads those page images and generates the
grounded answer.

\subsection{Ingestion}
\label{sec:ingestion}
\subsubsection{Priority-aware scheduling}
\label{sec:prio}
Three ingestion streams compete for the same GPU pool, fed to the
indexer as priority lanes: high for real-time analyst uploads on a
live deal, medium for hourly deal-document syncs, and low for slower
syncs of non-deal material. Analyst workflows must not block behind
either background stream. Scheduling splits across two layers: a
custom dispatcher chooses which \emph{document} to ingest next, and
vLLM~\cite{kwon2023efficient} chooses which \emph{embedding request}
the GPU runs next. Both preempt lower-priority work and resume evicted
jobs from the last durable page, so a burst of analyst uploads never
queues behind background syncs (App.~\ref{app:sched}).

\subsubsection{Skip-aware re-indexing}
\label{sec:skip}
The indexer skips redundant work at two levels (Fig.~\ref{fig:ingest}).
A SHA-256 of raw blob bytes gives a full-document skip. A SHA-256 of
each JPEG-encoded page image gives page-level skip, so re-indexing a
modified document re-embeds only changed pages. Orphan pages are
deleted when page count shrinks. The document hash is stamped only
after every page is durable in the vector index, so the retrieval path can filter
pages whose document is not yet stamped.

\subsubsection{Decoupled embedding and storage DPI}
\label{sec:dpi}
We render each page \emph{twice}: a 150~DPI render feeds ColQwen3-4B
for embeddings, and a separate 500~DPI render is stored and served to
the VLM at answer time. Both values come from ablations
(\S\ref{sec:dpi-ablation}, \S\ref{sec:hallu}). Rendering runs in a bounded
producer--consumer window, so the GPU stays fed while resident memory
is bounded by the window, not document length.

\subsection{Retrieval}
\label{sec:retrieval}
\begin{figure*}[t]
\centering
\includegraphics[width=\textwidth,keepaspectratio]{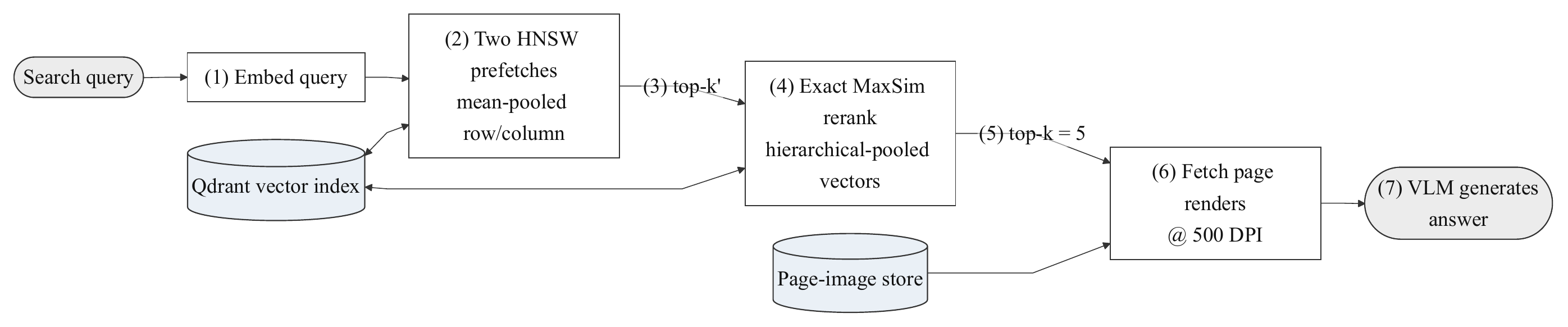}
\caption{PULSAR's retrieval path (\S\ref{sec:retrieval}): the query is
embedded with ColQwen3-4B, then two quantised
HNSW prefetches over mean-pooled row/column representations produce a
shortlist that exact MaxSim reranks over the hierarchical-pooled
vectors; the stored 500\,DPI render of each returned page grounds the
VLM.}
\label{fig:retrieval}
\end{figure*}
ColQwen3-4B emits a variable number of patch vectors per page,
each of dimension $D{=}320$. At firm scale (${\sim}2.4$M pages)
this is billions of vectors live in HNSW
\cite{malkov2020hnsw}, and exact MaxSim costs
$|q|{\times}|d|$ dot products per scored page. We therefore
cut that count by pooling~\cite{qdrant2026multivec_tut}, and keep exact MaxSim only on a short
rerank list \cite{santhanam2022plaid}.

\subsubsection{Pooling options}
\emph{Mean row/column pooling}
\cite{qdrant2026multivec_tut} treats the per-page patch sequence as
a 2D grid and averages along each axis, yielding row-
and column-mean vectors of dimension $D$. \emph{Hierarchical
pooling} \cite{clavie2024tokenpool} clusters similar patch vectors
and replaces each cluster by its mean; the \emph{pool factor} sets
the average cluster size, so pool factor~3 retains ${\sim}1/3$ of
the vectors, at which \citet{faysse2024colpali} report 97.8\% NDCG@5
retention. Mean gives a compact summary per page;
hierarchical keeps more detail at a higher per-page vector count.%
\footnote{Other variants binarise full-patch vectors with Hamming
MaxSim \cite{bergum2024vespacolpali}, or retrain the scoring rule
\cite{masry2025colmate}.} These pooling methods are known
individually, but no settled recipe combines them at scale, and the
usual two-stage index reranks on the full multivectors. PULSAR pools
that stage too, then sweeps the pooling-and-quantisation space for the
fastest configuration that holds retrieval quality (\S\ref{sec:lat}).

\subsubsection{Pooled two-stage retrieval}
The resulting index runs two pooled stages.

\textbf{Stage~1 (prefetch).} Each page is summarised
by 38--185 mean-pooled row/column vectors, binary-quantised in RAM
\cite{qdrant2026multivec_tut}. The query multivectors are left
unpooled; two HNSW prefetches against the row-pooled and column-pooled
page representations jointly produce a top-$k'$ shortlist
(Fig.~\ref{fig:retrieval}).

\textbf{Stage~2 (rerank).} The
shortlist is rescored with exact MaxSim over the hierarchical-pooled
multivectors~\cite{clavie2024tokenpool}, which sums, for each query
text token, its best dot product against any document
patch~\cite{khattab2020colbert}:
\begin{equation}
s(q,d) \;=\; \sum_{i=1}^{|q|} \max_{j\in[1,|d|]}\; q_i^{\!\top} d_j.
\label{eq:maxsim}
\end{equation}
The top-$k$ are returned. The pooled, quantised design reduces estimated
resident vector payload by 99.8\% and stored vector payload by 60.2\%
relative to the fully resident unpooled index (\S\ref{sec:footprint}).

\section{Experimental Setup}
\label{sec:setup}

The OCR+verbalisation baseline is the stack PULSAR replaced:
per-page OCR+layout extraction, figure verbalisation by the
query-time VLM (only filter-kept verbalisations are indexed;
\S\ref{sec:ingest}), semantic chunking, and dense embedding into
HNSW \cite{malkov2020hnsw} for dense+keyword hybrid retrieval.

\section{Results}
\label{sec:results}
We evaluate the retrieval index and its footprint
(\S\ref{sec:dpi-ablation}--\ref{sec:footprint}), ingestion cost and
refresh lag (\S\ref{sec:ingest}--\ref{sec:freshness}), then answer
quality (\S\ref{sec:hallu},
\S\ref{sec:headtohead}), where answer-fact recall more than
doubles over the OCR+verbalisation baseline.\footnote{In every experiment the queried index has its HNSW
graph verified fully built before measurement, so reported search
latencies are graph-based ANN, not a brute-force scan over unindexed
segments.}

\subsection{Embedding-DPI ablation}
\label{sec:dpi-ablation}
\begin{figure}[t]
\centering
\includegraphics[width=0.95\columnwidth]{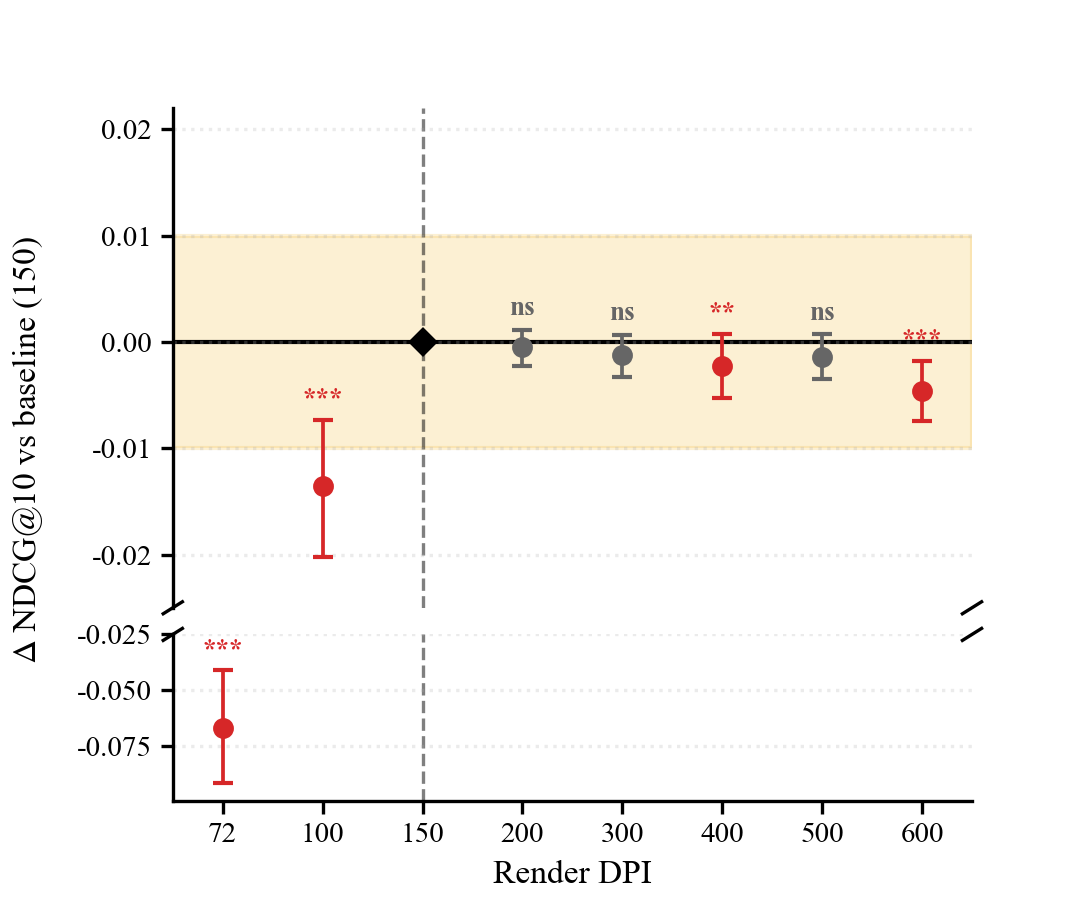}
\caption{Embedding-DPI ablation on ViDoRe~V3: mean paired $\Delta$NDCG@10 (vs.\ 150~DPI) against render DPI, 95\% bootstrap CIs, shaded $\pm0.01$ two one-sided tests equivalence band (\texttt{***}/\texttt{**}/\texttt{*}/\texttt{ns} significance).}
\label{fig:dpi_forest}
\end{figure}

Embedding DPI drives per-page GPU time and stored-vector count.
This GPU time dominates ingestion cost at ${\sim}2.4$M pages, so we
take the lowest DPI whose retrieval quality is statistically
equivalent to 150 (the a-priori baseline) and let ingestion cost
break the tie. We sweep DPI with a fixed-protocol ablation on
ViDoRe~V3 (single index,
identical query set).
\footnote{\label{fn:rule}ColQwen3-4B on ViDoRe~V3 ($n{=}14{,}514$
queries). Per-query paired Wilcoxon with Holm correction;
cluster-bootstrapped 95\% CIs over the eight ViDoRe~V3 sub-corpora. For the DPI
ablation we declare practical equivalence to 150 when the TOST 90\% CI
on the mean paired $\Delta$NDCG@10 falls inside $\pm0.01$ (NDCG points);
Cohen's $d$ (mean $\Delta$ over its paired SD) is reported as the
accompanying effect size.}
\footnote{Pages are re-rasterised from source PDFs at each target DPI.
ColQwen3's default
\texttt{max\_num\_visual\_tokens}{=}1280 caps input at ${\approx}1$\,MP
(${\approx}100$\,DPI on A4); we raise it to 65{,}536 so each render
is seen at full resolution. The plateau above 150~DPI is therefore a
property of ColQwen3 on these inputs.}
Across the seven DPIs compared with 150
(Fig.~\ref{fig:dpi_forest}), only 72 and 100 fall outside the $\pm0.01$
NDCG@10 band (Cohen's $d{=}{-}0.33$ and ${-}0.10$; both
$p_{\mathrm{Holm}}{<}10^{-29}$), so lower DPI costs measurable quality.
Every higher DPI (200--600) is equivalent to 150 (equivalence-test CI
inside $\pm0.01$; effect sizes $|d|{<}0.05$) but costs more per page to
ingest. PULSAR therefore uses 150~DPI.

\subsection{Pooling and quantisation ablation}
\label{sec:lat}
\begin{table}[t]
\centering\small
\setlength{\tabcolsep}{4pt}
\begin{tabular}{@{}lrrrr@{}}
\toprule
\textbf{Cfg} & $\Delta$\textbf{NDCG@10} & $\Delta$\textbf{Rec.} & \textbf{ms} & $\times$\\
\midrule
O (baseline)   & 0\phantom{$^{***}$}      & 0\phantom{$^{***}$}      & 752.7 & 1.0\\
O$_q$          & $+$.0004\phantom{$^{***}$} & $+$.0001\phantom{$^{***}$} & 439.4 & 1.7\\
H+O            & $-$.0002\phantom{$^{***}$} & $-$.0001\phantom{$^{***}$} & 429.8 & 1.8\\
H+O$_{qp}$     & $+$.0001\phantom{$^{***}$} & $-$.0002\phantom{$^{***}$} & 257.1 & 2.9\\
M+O            & $-$.0014\phantom{$^{***}$} & $-$.0023$^{**}$\phantom{$^{*}$} & 216.2 & 3.5\\
M+H            & $-$.0077$^{***}$ & $-$.0065$^{***}$ & 185.3 & 4.1\\
M+O$_{qp}$     & $-$.0026\phantom{$^{***}$} & $-$.0043$^{***}$ & 170.9 & 4.4\\
H              & $-$.0059$^{***}$ & $-$.0037$^{***}$ & 135.4 & 5.6\\
H$_q$          & $-$.0061$^{***}$ & $-$.0039$^{***}$ & \phantom{0}73.9 & 10.2\\
\textbf{M+H$_{qp}$} $\star$ & $-$.0088$^{***}$ & $-$.0074$^{***}$ & \textbf{\phantom{0}49.9} & \textbf{15.1}\\
\bottomrule
\end{tabular}
\caption{ViDoRe~V3 sweep over pooling and binary quantisation, ten
configurations. \texttt{O}/\texttt{M}/\texttt{H} are no/mean/hierarchical
pooling; \texttt{X+Y} is \texttt{X}-pooled prefetch, \texttt{Y}-pooled
rerank; $q$ marks in-place quantisation, $qp$ prefetch-only. $\Delta$
columns are paired differences vs.\ \texttt{O}; ms is median vector
search time; $\times$ is speedup over \texttt{O}.
\texttt{***}~$p_{\mathrm{Holm}}{<}.001$, \texttt{**}~${<}.01$; $\star$ is
production.}
\label{tab:index}
\end{table}
Lower vector search latency lets the same vector-database
capacity serve more concurrent searches at unchanged retrieval
quality, so we target the lowest latency that holds quality.
MaxSim scoring over full page multivectors is the dominant
per-query cost. We run a ten-configuration
sweep crossing pooling strategy with binary quantisation, taking the
faithful no-pooling no-quantisation \texttt{original} configuration as
the paired baseline and gating on a small effect size
($|d|{<}0.1$).\footnote{Unlike the DPI ablation
(\S\ref{sec:dpi-ablation}), this is not an equivalence question:
pooling and quantisation trade retrieval quality for cost, so rather
than certify parity we bound the acceptable loss by effect size on
both NDCG@10 and Recall@10, and let ingest or serving cost break the
tie.}

Table~\ref{tab:index} summarises the sweep. All nine non-baseline
configurations pass the quality gate, so the choice reduces to latency.
Latencies are medians over the same $n{=}14{,}514$ ViDoRe~V3
queries, each issued single-stream against a
single-replica vector database (Table~\ref{tab:hw}) holding all
ViDoRe~V3 pages.\footnote{Pages are indexed at 150~DPI; searches use
\texttt{hnsw\_ef\_search}${=}128$, with \texttt{prefetch\_limit}${=}200$
for the two-stage configurations.}
\texttt{M+H}$_{qp}$ (mean-pooled binary-quantised HNSW prefetch
with hierarchical-pooled rerank) gives the lowest median vector
search latency at 49.9\,ms (IQR 22.7\,ms), a $15.1\times$ speedup
over the 752.7\,ms baseline (IQR 472.7\,ms). The quality cost
($\Delta$NDCG@10 $=-0.0088$, $\Delta$Recall@10 $=-0.0074$; Cohen's
$d{=}{-}0.09$ and $-0.06$) clears the $|d|{<}0.1$ gate, so PULSAR runs
\texttt{M+H}$_{qp}$.

The speedup comes from two effects that compose. \emph{Pooling}
shrinks the per-page vector count the reranker scores; hierarchical
pooling alone (\texttt{H}) already gives $5.6\times$.
\emph{Binary quantisation} shrinks the per-comparison cost
(\texttt{O}$_q$: $1.7\times$). Quantised hierarchical rerank
(\texttt{H}$_q$) gives $10.2\times$.
\texttt{M+H}$_{qp}$ instead uses binary mean-pooled prefetch with
Float32 hierarchical reranking ($15.1\times$).\footnote{The prefetch only pays off once quantised:
unquantised (\texttt{M+H}, $4.1\times$) it is slower than hierarchical
rerank alone (\texttt{H}, $5.6\times$), because the prefetch HNSW lookup itself costs more than
it saves the reranker.}

On the chosen \texttt{M+H}$_{qp}$ index we also swept the two Qdrant
search knobs. \texttt{prefetch\_limit} rises in quality to a knee at
200 and is flat above; \texttt{hnsw\_ef\_search} has no measurable
quality effect across the swept range (Friedman $p{=}0.14$;
Table~\ref{tab:searchparams}). PULSAR serves \texttt{prefetch\_limit}{=}200
and \texttt{hnsw\_ef\_search}{=}64.

\subsection{Storage and latency costs}
\label{sec:footprint}
We validate the sweep's pooled two-stage choice
(\S\ref{sec:retrieval},~\S\ref{sec:lat}) against an unpooled index on
the dev slice (3{,}597 documents, 106{,}074 pages).
The pooled \texttt{M+H}$_{qp}$ index keeps binary mean-pooled prefetch
vectors pinned in RAM, with Float32 means and hierarchical rerank
vectors on disk. Estimated resident vector payload falls by 99.8\%
and stored vector payload by 60.2\% (Table~\ref{tab:footprint}).
At production scale, the binary prefetch vectors fit in cluster RAM,
whereas pinning the full unpooled page multivectors would exceed
the available capacity.

\begin{figure}[t]
\centering
\includegraphics[width=0.95\columnwidth]{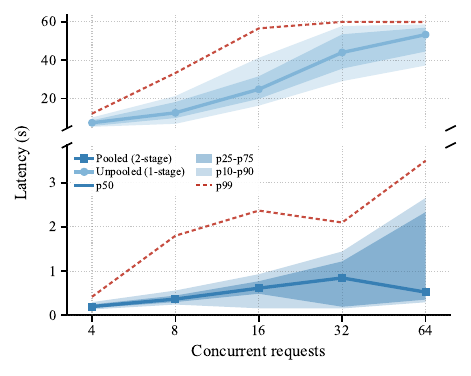}
\caption{Per-query latency vs.\ concurrent requests: p50 line,
p25--p75 and p10--p90 bands, p99 dashed. Pooled stays sub-second;
unpooled runs an order of magnitude higher. Throughput:
Fig.~\ref{fig:pooled-qps}.}
\label{fig:pooled-load}
\end{figure}

On a single query the pooled index answers in a median 175\,ms (95\% CI
165--186) against 1.9\,s for unpooled, a $10.9\times$ gap that widens
to $13\times$ at p95 and $20\times$ at p99.\footnote{Load tests run
over the 106{,}074-page dev slice on identical hardware. Both indexes
retrieve by HNSW ANN. The latency/throughput curve is single-client at 1--4-way and
distributed (8 pods) at 8--64-way, with completion/timeout accounting;
the single-query result uses fixed sample counts (pooled $n{=}300$,
unpooled $n{=}100$) after a 60\,s warmup, with bootstrap 95\% CIs. The
load test uses the dev Qdrant (Table~\ref{tab:hw}); production's stack
has ${\sim}2\times$ the Qdrant CPU and disk IOPS.}
As load rises the pooled index scales: its throughput climbs to
${\approx}47$ QPS at a sub-second median, every request completing. The
unpooled index does not: its throughput stays flat near 0.5 QPS while
latency runs to 53\,s, and it drops over half its requests at 64-way
(Figs.~\ref{fig:pooled-load} and~\ref{fig:pooled-qps}).
In production, the same pooled index (Table~\ref{tab:config}) holds a
156\,ms median vector-search latency over a rolling 7-day window
($n{=}109{,}370$ calls), with a concurrency-driven tail (p95 0.6\,s,
p99 1.5\,s) under a load that peaks at 62~QPS.

\subsection{Ingestion cost}
\label{sec:ingest}
On the dev slice, we estimate ingestion cost from call counts
priced at list rates, not from metered cloud spend. The
OCR+verbalisation baseline runs one OCR+layout call per page plus
one VLM call per figure; PULSAR runs one ColQwen3-4B forward pass
per page. On this basis the baseline's per-page
ingest cost is ${\approx}20\times$ PULSAR's.%
\footnote{The baseline sent 143{,}934 figures for verbalisation
across the slice; only 38{,}868 (27\%) yielded a description the
downstream filter kept. Every call is still paid for, so nearly
three-quarters of the per-figure VLM cost buys output that is
discarded.}

\subsection{Ingestion latency and refresh lag}
\label{sec:freshness}
Per-document end-to-end ingestion latency on the dev cluster
(Table~\ref{tab:hw}), measured across the dev slice, follows a rolling
median $T_{\mathrm{med}}(p) \approx 2.9 + 1.41\,p$ seconds
(Fig.~\ref{fig:ingest_perf}): a ${\sim}3$\,s
steady-state per-document overhead (object fetch, document open, index
upsert) plus ${\approx}1.4$\,s per page.

Freshness comes from skip-aware ingestion (\S\ref{sec:skip}): each
sync re-embeds only the pages that changed, not the whole
corpus, so refresh lag tracks edited-page count, not document length.
A single-page edit to a multi-hundred-page deck is searchable in
seconds. Over the 7-day window the pipeline cleared 50{,}368 changed
pages, but the change rate is bursty: the busiest hour alone cleared
5{,}491. Change-to-searchable lag had a median of 2.0\,s and a p95 of
43\,s; in that busiest hour p95 held at 63\,s and the worst case at
328\,s. Priority scheduling (\S\ref{sec:prio}) runs live-deal uploads
continuously and background deal material on an hourly cadence, keeping
analyst uploads ahead of the background refresh.

\subsection{Answer-time DPI ablation}
\label{sec:hallu}
With embedding DPI locked at 150 for retrieval, we run an
answer-time DPI ablation on two firm investment-committee decks
(516 questions in total; Table~\ref{tab:hallu}). Both are firm-authored IC packs with
the chart, table, and figure density typical of production
traffic. Only the page image served to the VLM at answer time
varied ($\{150, 400, 500\}$ DPI). For each page, a large
language model (LLM) generated three questions on its chart, table,
and figure content; SMEs validated the questions and authored the
gold answers. The same question set was reused at every DPI. SMEs labelled
each system answer as correct or as a page-legibility
error: an unreadable page image at that DPI,
not generation drift on a faithful page
\cite{ji2023survey}.

Page-legibility errors fall monotonically with DPI on both decks,
reaching zero at 500, so PULSAR serves 500~DPI at answer
time.\footnote{The 400~DPI residuals concentrate on dense charts with
small axes, grouped series, and tight legend-to-mark correspondences.
This is a controlled ablation over one variable on representative IC
pages, not a corpus-scale legibility benchmark.}

\subsection{PULSAR vs the baseline}
\label{sec:headtohead}
\begin{table*}[!b]
\centering\small
\setlength{\tabcolsep}{5pt}
\begin{tabular}{@{}l rrrr@{}}
\toprule
\textbf{Metric} & $k{=}1$ & $k{=}3$ & $k{=}5$ & $k{=}10$\\
\midrule
\multicolumn{5}{@{}l}{\textit{OCR+verbalisation baseline}}\\
Completeness (1--5) & 1.68\,$\pm$.02 & 2.34\,$\pm$.04 & 2.55\,$\pm$.03 & 2.85\,$\pm$.04\\
ContextFactRecall   & .134\,$\pm$.005 & .322\,$\pm$.004 & .388\,$\pm$.013 & .507\,$\pm$.008\\
AnswerFactRecall    & .102\,$\pm$.002 & .226\,$\pm$.010 & .271\,$\pm$.005 & .355\,$\pm$.003\\
\midrule
\multicolumn{5}{@{}l}{\textit{PULSAR (vision-first)}}\\
Completeness (1--5) & \textbf{3.41\,$\pm$.06} & \textbf{3.89\,$\pm$.07} & \textbf{4.15\,$\pm$.06} & \textbf{4.14\,$\pm$.04}\\
ContextFactRecall   & \textbf{.587\,$\pm$.004} & \textbf{.745\,$\pm$.006} & \textbf{.874\,$\pm$.006} & \textbf{.949\,$\pm$.002}\\
AnswerFactRecall    & \textbf{.492\,$\pm$.007} & \textbf{.650\,$\pm$.021} & \textbf{.733\,$\pm$.009} & \textbf{.729\,$\pm$.015}\\
\midrule
\multicolumn{5}{@{}l}{\textit{$\Delta$ deal-weighted (PULSAR $-$ baseline), 95\% CI over 7 deals}}\\
Completeness (1--5) & $+$1.90\,[1.18,\,2.62] & $+$1.75\,[1.26,\,2.24] & $+$1.77\,[1.28,\,2.25] & $+$1.51\,[0.78,\,2.25]\\
ContextFactRecall   & $+$0.51\,[0.35,\,0.66] & $+$0.47\,[0.28,\,0.66] & $+$0.55\,[0.37,\,0.73] & $+$0.50\,[0.29,\,0.72]\\
AnswerFactRecall    & $+$0.43\,[0.27,\,0.59] & $+$0.47\,[0.33,\,0.62] & $+$0.52\,[0.37,\,0.66] & $+$0.44\,[0.25,\,0.64]\\
\bottomrule
\end{tabular}
\caption{PULSAR vs the OCR+verbalisation baseline on $n{=}75$ SME-authored questions (\S\ref{sec:headtohead}); cells are mean\,$\pm$\,std over 5 passes. Completeness is a 1--5 LLM judgement of the system's answer against the gold answer; Context/AnswerFactRecall are the shares of gold facts in the retrieved context and in the answer. Bold marks PULSAR higher at the same $k$.}
\label{tab:headtohead}
\end{table*}
We compare PULSAR end to end against the OCR+verbalisation baseline
on answer quality. PULSAR scores higher on
Completeness and both context- and answer-fact recall at every
$k \in \{1,3,5,10\}$
(Table~\ref{tab:headtohead}). The evaluation set is $n{=}75$
SME-authored investment questions with SME-authored gold answers
(categorised in App.~\ref{app:qtax}); both retrieval paths run the
same questions and generate answers with the same answer-time VLM, so
only the indexing path differs (verbalised text chunks vs.\ page
images). An LLM judge (validated against blinded SME re-scoring;
App.~\ref{app:human})
scores each system's output against the gold. The two paths index
different units, so we report answer-level, modality-independent
metrics, not retrieval rank.

To separate run-to-run model noise from sampling uncertainty, we
cluster by deal (the correlated
unit) and report in the bottom block of Table~\ref{tab:headtohead} the
deal-weighted PULSAR$-$baseline delta with a 95\% bootstrap CI over the
7 deals.\footnote{The $\pm$ in Table~\ref{tab:headtohead} is
temperature-0 run-to-run non-determinism in the answer-time VLM and
LLM judge, not sampling error.} Every interval excludes 0 at every $k$; PULSAR beats the
baseline on all seven deals in all twelve metric-by-$k$ cells (exact
Wilcoxon signed-rank $p{=}.016$, the floor at $n{=}7$), and the
advantage survives leave-one-deal-out.

The advantage is retrieval-led. ContextFactRecall is $1.9$--$2.3\times$ the baseline
from $k{=}3$ up, and larger still at $k{=}1$, where the baseline
retrieves almost nothing (Table~\ref{tab:headtohead}). This
is consistent with vision-first retrieval preserving the chart
values and legend assignments condensed during
verbalisation. AnswerFactRecall inherits this gap: it is widest at
low $k$, where PULSAR reaches roughly $5\times$ the baseline
($.49$ vs $.10$ at $k{=}1$), and narrows as the baseline catches up.
At the production $k{=}5$
(\S\ref{sec:architecture}), PULSAR already exceeds the baseline's $k{=}10$
on every metric.

\section{Conclusion}
PULSAR runs vision-first RAG in production at firm scale. Two choices
keep it affordable: decoupling embedding DPI (150) from answer-time DPI
(500), and a pooled two-stage late-interaction index that cuts median
vector-search latency $15.1\times$ and sustains ${\approx}88\times$ an unpooled index's
throughput under load, at under 0.01 NDCG@10 and Recall@10
loss. Ingestion is ${\approx}20\times$
cheaper per page than the OCR+verbalisation baseline, and
answer-fact recall more than doubles.

\section*{Limitations}

\tightpara{Late interaction stays heavier than single-vector
retrieval.} Even with hierarchical pooling and the two-stage index
(\S\ref{sec:retrieval}), each page is still stored and scored as many
pooled vectors rather than one, and MaxSim remains costlier than a
single dot product. These optimisations cut late interaction's footprint
and scoring cost but do not erase this overhead: it stays heavier than
single-vector retrieval.

\tightpara{Findings are tied to this corpus, baseline, and
backbone.} All production measurements come from one firm's chart-
and table-heavy private-markets deal documents, run on ColQwen3-4B. The
DPI sweet spot, ingestion-cost gap, and factuality margins may shift
on corpora with different chart density or page complexity, on other
vision-first backbones, and against text-extraction stacks that
preserve more chart detail than the OCR+verbalisation baseline here.
Our retrieval-quality evidence for the index design, the embedding-DPI
and pooling/quantisation experiments
(\S\ref{sec:dpi-ablation},~\S\ref{sec:lat}), is measured on the public
ViDoRe~V3 benchmark rather than the firm corpus; we have not confirmed
these findings hold on the firm's own documents.

\tightpara{Ingest-cost estimate is ingestion-only.} The
${\approx}20\times$ gap (\S\ref{sec:ingest}) covers ingestion calls
only; standing serving and answer-time inference are excluded.

\tightpara{Answer-time DPI ablation uses SME ground truth and
LLM-generated questions.} The 516-question DPI result
(\S\ref{sec:hallu}) checks each DPI's answers against factual ground
truth the finance SMEs authored ahead of time. Labels are
value-level matches (e.g.\ does the answer's stated internal rate
of return equal the
SME's), not subjective quality judgements. The question-generating
LLM is in the same model family as the answer-time VLM, so both may
share the same visual blind spots. The protocol isolates the
answer-generation step: each question is generated from a specific
page, which is fed directly to the VLM, so retrieval is not exercised. It does not speak to
end-to-end behaviour, where retrieval can also fail. The 516 questions cluster on two decks, so the effective
sample size for between-deck generalisation is two; the result
speaks to behaviour on IC-pack-style pages, not arbitrary documents.

\tightpara{We compare against one baseline, not alternative
vision-first stacks.} The OCR+verbalisation baseline vs vision-first comparison in
\S\ref{sec:headtohead} runs 75 SME-authored questions through
both systems and scores three LLM-judged metrics against
SME-authored gold answers, repeated over 5 independent passes. A
blind 20\% subsample re-scored by three domain experts tracks the
judge (weighted $\kappa\ge0.79$) and independently favours PULSAR
(App.~\ref{app:human}), so the comparison is human-validated, not
judge-only. We do not compare against alternative
vision-first stacks (e.g.\ OCR-free document transformers
\cite{kim2022donut,hu2024docowl}) or generate-and-encode hybrids
\cite{nguyen2025serval}, so we do not claim late-interaction
page-image retrieval is the best approach for this task.

\section*{Ethical Considerations}

\tightpara{Decision impact.} PULSAR informs investment decisions, so
even a low residual error
rate carries downstream consequences. It is surfaced to analysts as
an advisory aid, and outputs that propagate into client-facing
artefacts go through human review. We do not recommend deploying
vision-first RAG as an unsupervised decision-maker on investment
material.

\tightpara{Data.} Retrieval quality is benchmarked on the public
ViDoRe~V3 suite. The production deployment, the end-to-end answer
evaluation, and the load and ingestion measurements run on Mubadala's
confidential deal documents, which are not released and are not used
to train or fine-tune any model: PULSAR runs a frozen pretrained
backbone, and all indexing and evaluation of that corpus happen within
the firm's own environment under its data-governance controls. The
evaluation questions and gold answers were authored by SMEs already
authorised to access the underlying documents.

\section*{Acknowledgements}
We thank colleagues at Microsoft, Inception42, and Mubadala for their
internal review of the paper and their help in evaluating outputs, and
the anonymous reviewers for their feedback.

\bibliography{refs}

\appendix

\setcounter{topnumber}{4}
\setcounter{bottomnumber}{3}
\setcounter{totalnumber}{6}
\renewcommand{\topfraction}{0.95}
\renewcommand{\bottomfraction}{0.95}
\renewcommand{\textfraction}{0.05}
\renewcommand{\floatpagefraction}{0.6}

\section{Configuration and Hardware Setup}
\label{app:config}
Table~\ref{tab:config} lists the index and serving configuration and
Table~\ref{tab:hw} the Kubernetes cluster sizing, each across the
benchmark, dev, and prod settings. The benchmark
columns give the settings behind the \S\ref{sec:dpi-ablation} and
\S\ref{sec:lat} ablations; the dev and prod columns describe the
deployed system.

\begin{table}[h]
\centering\small
\setlength{\tabcolsep}{4pt}
\resizebox{\columnwidth}{!}{%
\begin{tabular}{@{}llll@{}}
\toprule
\textbf{Parameter} & \textbf{Benchmark} & \textbf{Dev} & \textbf{Prod}\\
\midrule
Backbone & \multicolumn{3}{l}{ColQwen3-4B (Qwen3-VL backbone)}\\
Model dtype & \multicolumn{3}{l}{bfloat16}\\
Vector dim $D$ & \multicolumn{3}{l}{320}\\
\texttt{patch\_size} / \texttt{merge\_size} & \multicolumn{3}{l}{16 / 2}\\
Distance metric & \multicolumn{3}{l}{cosine (MaxSim multivector)}\\
\texttt{max\_num\_visual\_tokens} & 65{,}536 & \multicolumn{2}{l}{4{,}096}\\
Image cap \texttt{max\_pixels} & 67{,}108{,}864 & \multicolumn{2}{l}{4{,}194{,}304}\\
Embedding/index DPI & swept & \multicolumn{2}{l}{150}\\
Stored/served DPI & n/a & \multicolumn{2}{l}{500}\\
\midrule
Vector DB engine & Qdrant 1.16 & \multicolumn{2}{l}{Qdrant 1.17}\\
HNSW graph degree $m$ & \multicolumn{3}{l}{16}\\
HNSW \texttt{ef\_construct} & \multicolumn{3}{l}{100}\\
Index configuration & swept & \multicolumn{2}{l}{\texttt{M+H}$_{qp}$}\\
Prefetch pooling & swept & \multicolumn{2}{l}{mean row + column}\\
Rerank pooling & swept & \multicolumn{2}{l}{hierarchical, pool factor 3}\\
Quantisation & swept & \multicolumn{2}{l}{binary, prefetch-only}\\
\texttt{hnsw\_ef\_search} & swept & \multicolumn{2}{l}{64}\\
\texttt{prefetch\_limit} & swept & \multicolumn{2}{l}{200}\\
\texttt{exact} & \multicolumn{3}{l}{false}\\
\texttt{indexed\_only} & false & \multicolumn{2}{l}{true}\\
\midrule
Serving engine & transformers & \multicolumn{2}{l}{vLLM 0.19.1}\\
\texttt{max\_model\_len} & n/a & \multicolumn{2}{l}{3{,}072}\\
GPU KV cache size & n/a & \multicolumn{2}{l}{4{,}560 tokens}\\
\texttt{cpu\_offload\_gb} & n/a & \multicolumn{2}{l}{0.3}\\
KV cache dtype & n/a & \multicolumn{2}{l}{bf16}\\
\texttt{gpu\_memory\_utilization} & n/a & \multicolumn{2}{l}{0.89}\\
\texttt{enforce\_eager} & n/a & \multicolumn{2}{l}{true}\\
\texttt{max\_num\_batched\_tokens} & n/a & \multicolumn{2}{l}{8{,}192}\\
\texttt{max\_num\_seqs} & n/a & \multicolumn{2}{l}{16}\\
\midrule
Returned pages & @10 (eval) & \multicolumn{2}{l}{top-$k{=}5$}\\
Query issue & single-stream & \multicolumn{2}{l}{concurrent}\\
\bottomrule
\end{tabular}}
\caption{Index/serving configuration for the offline benchmark and the
deployed system (dev and prod, both full-res A4 via CPU weight
offload).\protect\footnotemark{} \emph{swept} marks parameters varied in the
\S\ref{sec:dpi-ablation} and \S\ref{sec:lat} ablations; cluster sizing
is in Table~\ref{tab:hw}. Binary quantisation rescores at $2{\times}$
oversampling.}
\label{tab:config}
\end{table}
\footnotetext{The pixel (\texttt{max\_pixels}) and sequence-length
(\texttt{max\_model\_len}) caps are set so a standard A4 page embeds
without clipping.}

\begin{table}[h]
\centering\small
\setlength{\tabcolsep}{4pt}
\resizebox{\columnwidth}{!}{%
\begin{tabular}{@{}llll@{}}
\toprule
\textbf{Component} & \textbf{Benchmark} & \textbf{Dev} & \textbf{Prod}\\
\midrule
Transformers inference & $1{\times}$ (1/4,\,8/16) & -- & --\\
vLLM inference & -- & $2{\times}$ (2/4,\,12/20) & $4{\times}$ (2/4,\,16/24)\\
GPU node              & \multicolumn{3}{l}{\texttt{NV18ads\_A10\_v5} (all)}\\
\texttt{colpali-inference} & --          & $2{\times}$ (1/4,\,6/12)   & $4{\times}$ (1/4,\,6/12)\\
Doc indexer           & --          & $2{\times}$ (1/2,\,6/24)   & $4{\times}$ (1/2,\,6/24)\\
Vector DB             & $1{\times}$ (6/7.5,\,25/28) & $3{\times}$ (6/10,\,50/100) & $4{\times}$ (10/14,\,64/100)\\
Vector DB PVC         & 512\,GiB (2{,}300) & 512\,GiB P20 (2{,}300) & 1\,TiB P30 (5{,}000)\\
\bottomrule
\end{tabular}}
\caption{Kubernetes sizing (replicas~$\times$~CPU/mem, requests/limits),
persistent-volume managed-disk tiers with base provisioned IOPS in
parentheses; each inference pod gets one \texttt{NV18ads\_A10\_v5}
(a fractional-A10 SKU), so the GPU count equals the inference replica
count. The benchmark ran as
Azure ML pipeline jobs (AKS-backed) on a single-replica stack
(512\,GiB \texttt{managed-premium} Qdrant PVC): GPU work on
\texttt{NV18ads\_A10\_v5}, CPU work on \texttt{Standard\_D8s\_v3}. It
uses no separate doc-indexer component (--).}
\label{tab:hw}
\end{table}

\section{Priority-aware ingestion scheduling}
\label{app:sched}
Scheduling (\S\ref{sec:prio}) splits across two layers.

\tightpara{Document-level (dispatcher).} The dispatcher pulls from the
highest non-empty lane and a higher-priority arrival evicts any
in-flight lower-priority job. Each page is stamped with its page hash
at upsert (not the document hash, which is stamped only once the whole
document is done, \S\ref{sec:skip}), so on redelivery the evicted job resumes from the last
durable page; wasted work is bounded by the in-flight batch, not
document length.

\tightpara{Request-level (vLLM).} Within a lane, each embedding
request is stamped with a numeric priority derived from remaining
pages, so the GPU drains shortest remaining processing time (SRPT)
first and short documents do not queue behind long ones. A
time-in-flight term ages low-priority work up so it cannot starve.
Search queries enter at top priority, so foreground search is never
blocked by background embedding.

\section{Search-parameter Sweeps}
\label{app:searchparams}
Table~\ref{tab:searchparams} gives the full \texttt{prefetch\_limit}
and \texttt{hnsw\_ef\_search} sweeps behind the search-knob choices in
\S\ref{sec:lat}.

\begin{table}[h]
\centering\small
\setlength{\tabcolsep}{4pt}
\begin{tabular}{@{}lrrrrr@{}}
\toprule
\textbf{Setting} & $\Delta$\textbf{NDCG@10} & $p_{50}$ & $p_{90}$ & $p_{95}$ & $p_{99}$\\
\midrule
\multicolumn{6}{@{}l}{\textit{\texttt{prefetch\_limit} (baseline 200)}}\\
25                     & $-$.032$^{***}$ & 36 & \phantom{0}57 & \phantom{0}66 & \phantom{0}85\\
50                     & $-$.016$^{***}$ & 38 & \phantom{0}59 & \phantom{0}68 & \phantom{0}88\\
100                    & $-$.006$^{***}$ & 43 & \phantom{0}66 & \phantom{0}76 & 105\\
150                    & $-$.002$^{***}$ & 46 & \phantom{0}71 & \phantom{0}81 & 105\\
\textbf{200}$^{\star}$ & 0\phantom{$^{***}$} & 50 & \phantom{0}77 & \phantom{0}88 & 110\\
250                    & ns\phantom{$^{***}$} & 54 & \phantom{0}81 & \phantom{0}93 & 121\\
300                    & ns\phantom{$^{***}$} & 57 & \phantom{0}86 & \phantom{0}97 & 123\\
400                    & ns\phantom{$^{***}$} & 62 & \phantom{0}92 & 104 & 131\\
800                    & ns\phantom{$^{***}$} & 87 & 133 & 153 & 193\\
\midrule
\multicolumn{6}{@{}l}{\textit{\texttt{hnsw\_ef\_search} (baseline 128)}}\\
\textbf{64}$^{\star}$  & ns\phantom{$^{***}$} & 50 & \phantom{0}79 & \phantom{0}90 & 113\\
128                    & 0\phantom{$^{***}$} & 53 & \phantom{0}82 & \phantom{0}94 & 119\\
256                    & ns\phantom{$^{***}$} & 53 & \phantom{0}82 & \phantom{0}94 & 120\\
512                    & ns\phantom{$^{***}$} & 54 & \phantom{0}81 & \phantom{0}91 & 113\\
\bottomrule
\end{tabular}
\caption{Search-parameter sweeps on ViDoRe~V3 (\S\ref{sec:lat}), on
the selected \texttt{M+H}$_{qp}$ index at 150~DPI, $n{=}14{,}514$
queries. $\Delta$NDCG@10 is the mean paired difference vs.\ each
sweep's baseline (Wilcoxon + Holm within sweep; \emph{ns} =
indistinguishable on NDCG@10). $p_{50}$--$p_{99}$ are per-query Qdrant
server-side latencies (ms). \texttt{***}~$p_{\mathrm{Holm}}{<}.001$.
$\star$ marks the deployed setting.}
\label{tab:searchparams}
\end{table}

\section{Index Footprint}
\label{app:footprint}
Table~\ref{tab:footprint} gives estimated vector payloads for
106{,}074 pages, summarised in \S\ref{sec:footprint}.

\begin{table}[h]
\centering\small
\setlength{\tabcolsep}{3pt}
\begin{tabular}{@{}lrrr@{}}
\toprule
 & \textbf{Pooled} & \textbf{Unpooled} & \\
 & (\texttt{M+H}$_{qp}$) & (\texttt{O}) & $\times$ smaller\\
\midrule
Resident vector payload & 0.535\,GB & 227.72\,GB & 425.3\\
Stored vector payload   & 90.67\,GB & 227.72\,GB & 2.51\\
\bottomrule
\end{tabular}
\caption{Estimated vector payload for 106{,}074 pages, per logical copy,
in decimal GB. Pooled pins binary mean-pooled prefetch values in RAM
and stores Float32 means and hierarchical rerank vectors on disk,
along with binary copies of the means. The unpooled configuration
compared here keeps every full per-page Float32 multivector resident
and persisted. Figures exclude non-vector structures and cache residency
and are not total serving-RAM requirements.}
\label{tab:footprint}
\end{table}

\section{Retrieval Throughput Under Load}
\label{app:throughput}
\begin{figure}[h]
\centering
\includegraphics[width=\columnwidth]{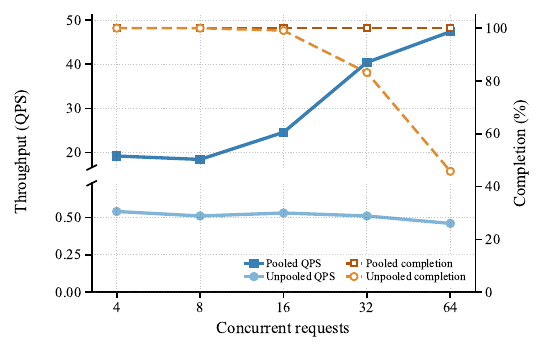}
\caption{Throughput (left, QPS) and request completion (right)
against concurrent requests, 106{,}074-page dev slice. Pooled climbs
to ${\approx}47$ QPS and holds 100\% completion; the unpooled index
tops out at 0.54 QPS (${\approx}88\times$ lower)
and its completion collapses under load, from 100\% to 46\% at 64-way, because each
unpooled search is so costly the system saturates at one-way
concurrency. Per-query latency in Fig.~\ref{fig:pooled-load}.}
\label{fig:pooled-qps}
\end{figure}

Fig.~\ref{fig:pooled-qps} reports the throughput behind the latency
comparison in \S\ref{sec:footprint}: throughput over a 4--64-way sweep
for both indexes at corpus scale.

We also isolate per-query latency from queueing with a single-concurrency
run (Table~\ref{tab:conc1-latency}): warm, no concurrent load, fixed sample
counts, bootstrap 95\% CIs. The pooled index answers in a median 175\,ms
against 1.9\,s for unpooled, and the gap widens through the tail.

\begin{table}[h]
\centering\small
\setlength{\tabcolsep}{5pt}
\begin{tabular}{@{}lrrrr@{}}
\toprule
 & \textbf{p50} & \textbf{p90} & \textbf{p95} & \textbf{p99}\\
\midrule
Pooled   & 175 & 387 & 546 & 686\\
Unpooled & 1{,}907 & 4{,}628 & 6{,}966 & 13{,}639\\
\bottomrule
\end{tabular}
\caption{Single-query latency (ms) at concurrency 1 over the
106{,}074-page dev slice, warmup discarded (pooled $n{=}300$, unpooled
$n{=}100$). Bootstrap 95\% CIs: pooled p50 165--186, p95 405--621;
unpooled p50 1{,}860--1{,}966, p95 4{,}351--8{,}748.}
\label{tab:conc1-latency}
\end{table}

\section{Ingestion Latency}
\label{app:ingestlat}
\begin{figure}[h]
\centering
\includegraphics[width=\columnwidth]{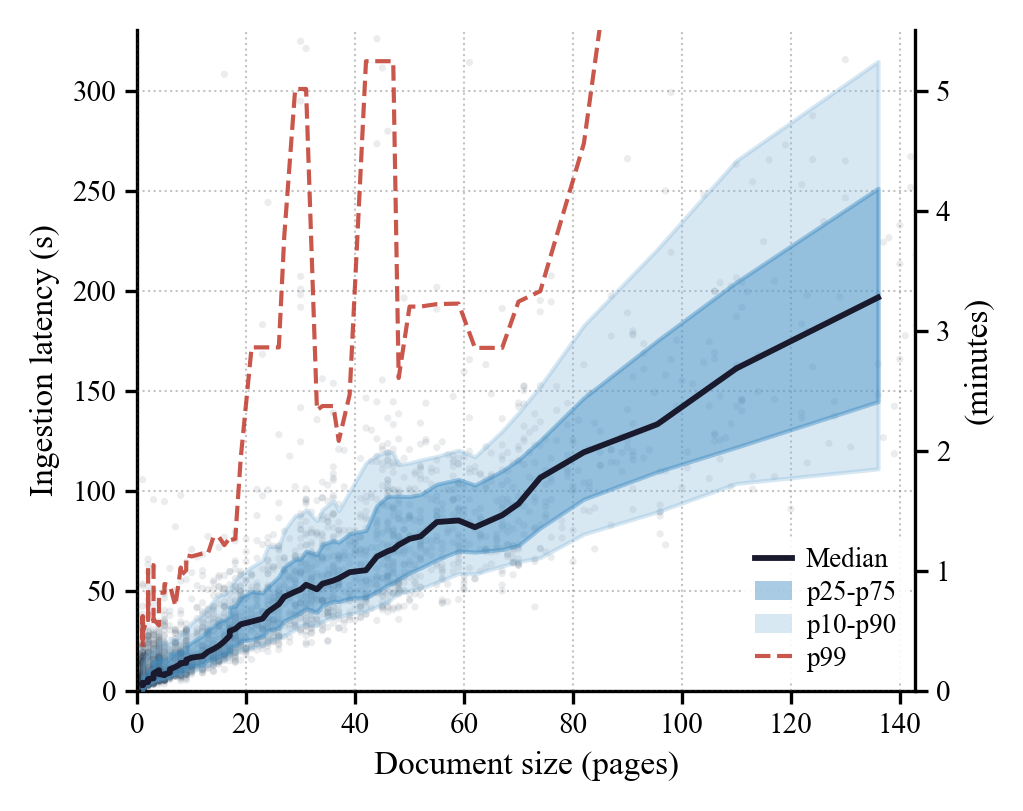}
\caption{End-to-end ingestion latency vs.\ document length
(dev cluster, 2 indexer pods $\times$ concurrency~2): rolling median,
IQR, p10--p90, and p99; x-axis clipped at
the 97th percentile (143 pages).}
\label{fig:ingest_perf}
\end{figure}

Fig.~\ref{fig:ingest_perf} plots the per-document ingestion latency
summarised in \S\ref{sec:freshness}, on the dev slice.

\section{Answer-time DPI Legibility Errors}
\label{app:hallu}
Table~\ref{tab:hallu} gives the per-deck page-legibility error counts
behind the answer-time DPI ablation in \S\ref{sec:hallu}.

\begin{table}[h]
\centering\small
\setlength{\tabcolsep}{4pt}
\begin{tabular}{@{}lrrccc@{}}
\toprule
\textbf{Deck} & \textbf{Pages} & \textbf{Qs} & \textbf{Vis. 150} & \textbf{Vis. 400} & \textbf{Vis. 500}\\
\midrule
Deck A & 94 & 282 & 18 & 8 & \textbf{0}\\
Deck B & 78 & 234 & 21 & 7 & \textbf{0}\\
\bottomrule
\end{tabular}
\caption{Page-legibility errors on two firm investment-committee decks
as answer-time DPI varies (\S\ref{sec:hallu}).}
\label{tab:hallu}
\end{table}

\section{Evaluation Question Taxonomy}
\label{app:qtax}
We categorise the $n{=}75$ head-to-head evaluation questions
(\S\ref{sec:headtohead}) on two axes, assigning each question one
label per axis. Examples are real questions with company, deal, and
competitor names removed.

\tightpara{Financial subject.}
(i)~Operating and financial metrics (22; 29.3\%):
``What is the projected trajectory for EBITDA and margins over the
forecast period for the company?''
(ii)~Market and competitive context (20; 26.7\%):
``What does the slide indicate about the projected growth and CAGR of
the overall sector market over the forecast period?''
(iii)~Valuation and returns (14; 18.7\%):
``What overall company valuation is implied by the proposed entry
share price?''
(iv)~Capital structure and funding (12; 16.0\%):
``What is the aggregate amount and denomination of the rated senior
and junior debt facilities?''
(v)~Segment and revenue mix (7; 9.3\%):
``How is the mix of contracted and non-contracted revenue projected to
change over the forecast period?''

\tightpara{Question operation.}
(i)~Value extraction (54; 72.0\%):
``What is the reported daily active user count for the platform?''
(ii)~Cross-entity or benchmark comparison (13; 17.3\%):
``What are a peer's total payment volume and EBITDA margin, and how
does its profitability measure up against the company's?''
(iii)~Temporal classification, reported vs projected (8; 10.7\%):
``Which year does the company report actual net revenue and EBITDA
margin figures, rather than a forward projection?''

\section{Human Validation of the LLM Judge}
\label{app:human}
To check that the LLM-judged head-to-head (\S\ref{sec:headtohead})
reflects human judgement, three domain-expert SMEs blindly re-scored a
deal-stratified 20\% subsample (15 questions, 64 atomic gold facts) of
the pass-1 answers. The two systems were shown as ``Answer~A''/``Answer~B''
in randomised order with modality cues removed, so raters scored
content only, on the same two metrics (Completeness 1--5,
AnswerFactRecall 0/1 per fact).

\tightpara{Agreement with the judge.} Per rater, Completeness tracks the
judge at Pearson $r{=}0.83$--$0.86$ and quadratic-weighted
$\kappa{=}0.79$--$0.81$, and AnswerFactRecall at $r{=}0.94$--$0.98$
(Table~\ref{tab:human}). Agreement between raters is ``almost perfect''
(Landis--Koch): pairwise weighted $\kappa{=}0.84$--$0.96$ on
Completeness, and Fleiss $\kappa{=}0.89$ on AnswerFactRecall over 128
fact-answer judgements (64 facts $\times$ 2 systems).

\tightpara{Direction confirmed.} All three SMEs independently rank
PULSAR above the baseline, each by a margin at least as large as the
judge's (Completeness $+1.07$ to $+1.60$ vs the judge's $+1.07$;
AnswerFactRecall $+0.32$ to $+0.41$ vs $+0.32$), so the judge is
conservative about the gap rather than inflating it; where human and
judge diverge, the human penalises PULSAR's own hallucinations harder.

\begin{table}[h]
\centering\small
\setlength{\tabcolsep}{6pt}
\begin{tabular}{@{}lccc@{}}
\toprule
 & \multicolumn{2}{c}{\textbf{Completeness}} & \textbf{FactRec.}\\
\textbf{Rater} & $r$ & wt.\ $\kappa$ & $r$\\
\midrule
SME1 & 0.86 & 0.81 & 0.98\\
SME2 & 0.84 & 0.80 & 0.98\\
SME3 & 0.83 & 0.79 & 0.94\\
median rater & 0.85 & 0.81 & --\\
\bottomrule
\end{tabular}
\caption{Agreement of three blind SME raters with the LLM judge on the
20\% subsample (\S\ref{sec:headtohead}): Pearson $r$ and
quadratic-weighted $\kappa$ for Completeness, $r$ for AnswerFactRecall.
The \emph{median rater} correlates the per-answer median of the three
SMEs, taken as one consensus rating, against the judge.}
\label{tab:human}
\end{table}

\end{document}